\documentclass[epjCONF]{svjour}
\usepackage{graphics}
\usepackage[varg]{txfonts} 
\usepackage[latin1]{inputenc}

\newcommand{\pythia}{{\sc pythia}}
\newcommand{\powheg}{{\sc powheg}}
\newcommand{\powhegbox}{{\sc powheg-box}}
\newcommand{\powhegpyt}{{\sc powheg+pythia}}
\newcommand{\powhegher}{{\sc powheg+herwig}}
\newcommand{\herwig}{{\sc herwig}}
\newcommand{\jimmy}{{\sc jimmy}}
\newcommand{\herwigpp}{{\sc herwig{\small ++}}}
\newcommand{\alpgen}{{\sc alpgen}}
\newcommand{\hej}{{\sc hej}}

\usepackage{subfigure}
\session-title{Hadron Collider Physics Symposium 2011}
\begin{document}
\title{Measurement of dijet production with a veto on additional central jet activity in pp collisions at $\sqrt{s} = 7~TeV$ using the ATLAS detector}
\author{Pauline Bernat\thanks{\email{pauline.bernat@cern.ch}}}
\institute{University College London - Gower Street - London - WC1E 6BT\\
\textit{(On behalf of the ATLAS Collaboration})}

\abstract{
A measurement of jet activity in the rapidity interval bounded by a dijet system is performed using pp collisions at 7 TeV recorded by the ATLAS detector in 2010. The data are compared to  LO predictions from \pythia{},  \herwigpp{} and \alpgen{} event generators. The data are also compared to NLO parton shower  prediction from POWHEG, when interfaced to \pythia{} or \herwig{} parton shower, and all order resummation prediction from HEJ. In most of the phase-space regions presented, the experimental uncertainty is much smaller than the spread of LO Monte Carlo event generator predictions. In general, \powhegpyt{} gave the best description of the data. 
} 

%
\maketitle
\section{Motivations}
\label{intro}
In this measurement, the QCD activity, or the absence of additional jet activity in dijet production is studied using a jet veto. Diverse perturbative QCD effects between widely separated jets can be investigated, the veto scale being much larger than $\Lambda_{\rm QCD}$. First, BFKL-like dynamics are expected to become increasingly important for large rapidity intervals \cite{Mueller:1986ey,Andersen:2010ih}. Alternatively, the effects of wide-angle soft-gluon radiation can be studied in the limit that the average dijet transverse momentum is much larger than the scale used to veto on additional jet activity \cite{Forshaw:2009fz}. Finally, colour singlet exchange is expected to be important if both limits are satisfied at the same time, i.e the jets are widely separated and the jet veto scale is small in comparison to the dijet transverse momentum.
 This work aims at measuring and comparing the effects of QCD radiation to standard event generators in those regions of 
phase space explored for the first time~\cite{Simone}. This jet veto measurement is also very useful to constrain the theoretical modelling in the current search for Higgs production via vector boson fusion and future precision Higgs measurements \cite{Campbell:2006xx}. 
\section{Measurements and Event Selection}
\label{sec:0}
Jets are reconstructed using the anti-$k_{\rm t}$ algorithm \cite{Cacciari:2008gp} with distance parameter $R=0.6$ and full four momentum recombination. Jets are required to have transverse momentum $p_{\rm T}^{}>20$~GeV and rapidity $|y| <4.4$. The dijet system is identified using two different selection criteria. In the first, the dijet system is made of the two highest transverse momentum jets in the event, which probes wide-angle soft gluon radiation in $p_{\rm T}$-ordered jet configurations. In the second, the most forward and the most backward jets in the event are used, which favours BFKL-like dynamics because the dijet invariant mass is much larger than the transverse momentum of the jets. 
For both definitions, the mean transverse momentum of the jets that define the dijet system, $\overline{p}_{\rm T}$, is required to be greater than 50~GeV. 
The variable used to quantify the amount of additional radiation in the rapidity interval bounded by the dijet system is the gap fraction. This is the fraction of events with no additional jet with a transverse momentum greater than a given veto scale, $Q_{0}$, in the rapidity interval bounded by the dijet system\footnote{A second variable is the mean number of jets with $p^{}_{\rm T}>Q_{0}$ in the rapidity interval bounded by the dijet system. See \cite{our-paper} for related results.}. By default, $Q_{0}=20$~GeV. The measurements are fully corrected for experimental effects. The final distributions therefore correspond to the `hadron-level', in which the jets are reconstructed using all final state particles that have a proper lifetime longer than 10~ps. 

The measurement was performed using data taken during 2010. The primary trigger selections used to readout the ATLAS detector were the calorimeter jet triggers. In particular, distinct regions of $\overline{p}_{\rm T}$ were defined and, in each region, only events that passed a specific jet trigger (at least one jet above a defined threshold) were used.  To minimise the impact of pile-up, each event was required to have exactly one reconstructed primary vertex, defined as a vertex with at least five tracks that was consistent with the beamspot. The gap fraction was found to be independent of the data taking period after the single vertex requirement. 

\section{Corrections for detector effects}
\label{sec:2}
The corrections for detector effects were calculated with a bin-by-bin unfolding procedure. The correction applied on the data is defined in each bin as the ratio of the hadron-level distribution (including muons and neutrinos) to the detector level distributions, using the \pythia{} event generator. The typical correction factor was observed to be a few percent in the rapidity interval between the boundary jets. 

The systematic uncertainty on the detector correction was determined considering both the physics modelling uncertainty, accounting  for any deviation between data and Monte Carlo (MC) and JES uncertainty impact on shapes, and the detector modelling uncertainty, estimated by varying the jet reconstruction efficiency and the jet energy resolution within the allowed uncertainties as determined from the data, and reweighting the $z$-vertex distribution. The total systematic uncertainty in the detector correction was defined as the quadrature sum of the physics/detector modelling uncertainties and the statistical uncertainty of the \pythia{} samples. The systematic uncertainty is typically 2-3\% and increases with $\Delta y$ and $\overline{p}_{\rm T}$, due to larger statistical uncertainty in the MC samples. 

 The total uncertainty due to systematic effects is the sum in quadrature of the uncertainty due to JES and the uncertainty due to the correction for detector effects.

\section{Data comparison to Leading Order event generators}
\label{sec:3}

Data are compared to different MC event generators: \pythia{} 6.4.2.3 \cite{pythia}, the reference generator,  \herwigpp{} 2.5.0 \cite{Bahr:2008pv} and \alpgen{} \cite{ref:noteALPGEN}. These fully simulated event samples are also used to derive systematic uncertainties and correction factor for detector effects. 
 The events were generated using the MRST LO* parton distribution functions (PDF)  \cite{Martin:2009iq}  in \pythia{} and \herwigpp{} and using, respectively, the AMBT1 tune~\cite{ATLAS-CONF-2010-031} and the LHC-UE7-1 tune \cite{Gieseke:2011na}. The \alpgen{} samples were generated using the CTEQ6L1 PDF set  \cite{Pumplin:2002vw} and passed through \herwig{} 6.5 \cite{herwig} and \jimmy{} \cite{Butterworth:1996zw} to provide parton showering, hadronisation and multiple partonic interactions (and the tune AUET1  \cite{auet}).

\begin{figure}[h]
\vspace{-0.4cm} 
 \hspace{2cm} (a) \hspace{3.5cm} (b)\\
\resizebox{1.\columnwidth}{!}{

\includegraphics{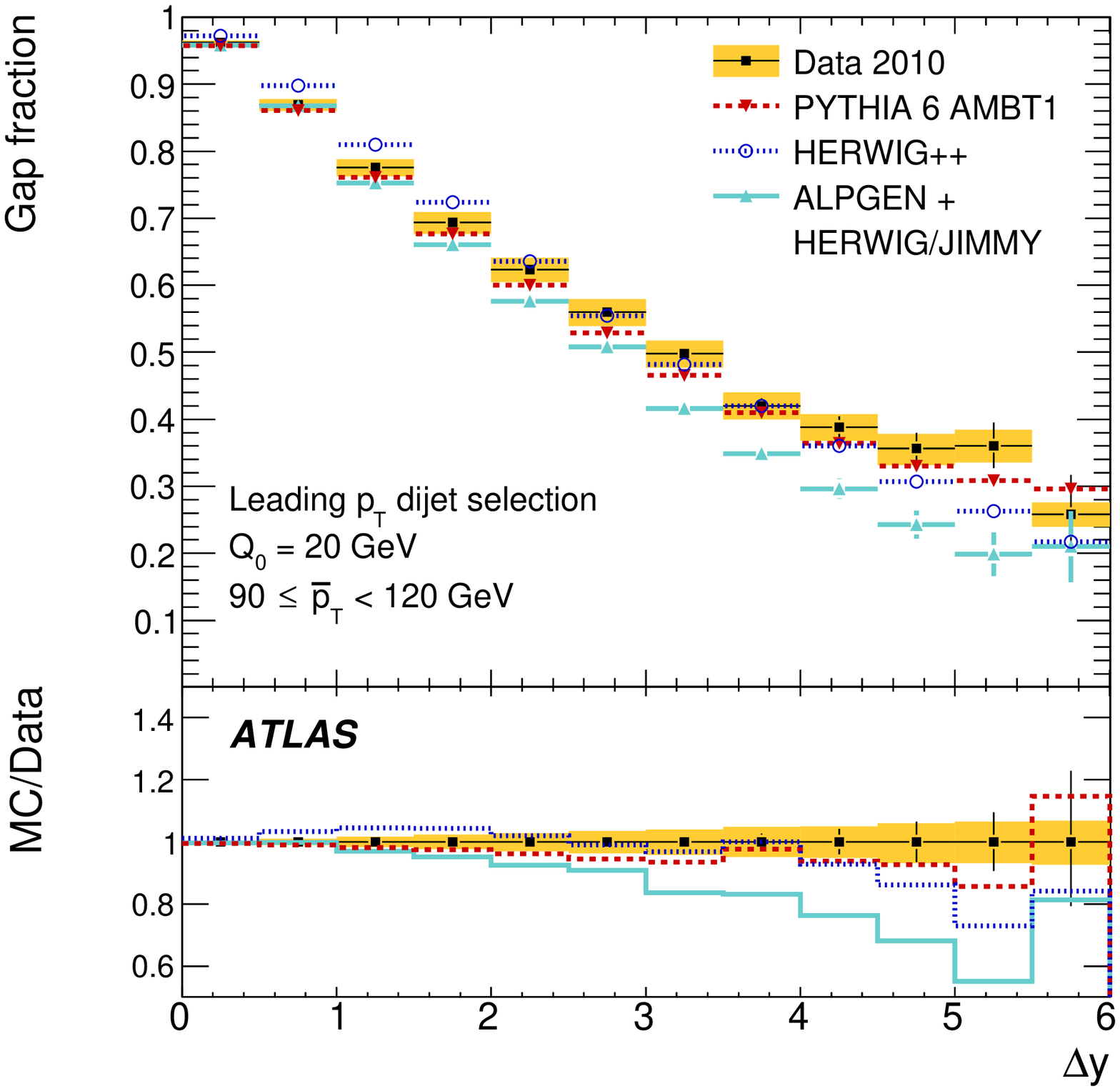}\\
\includegraphics{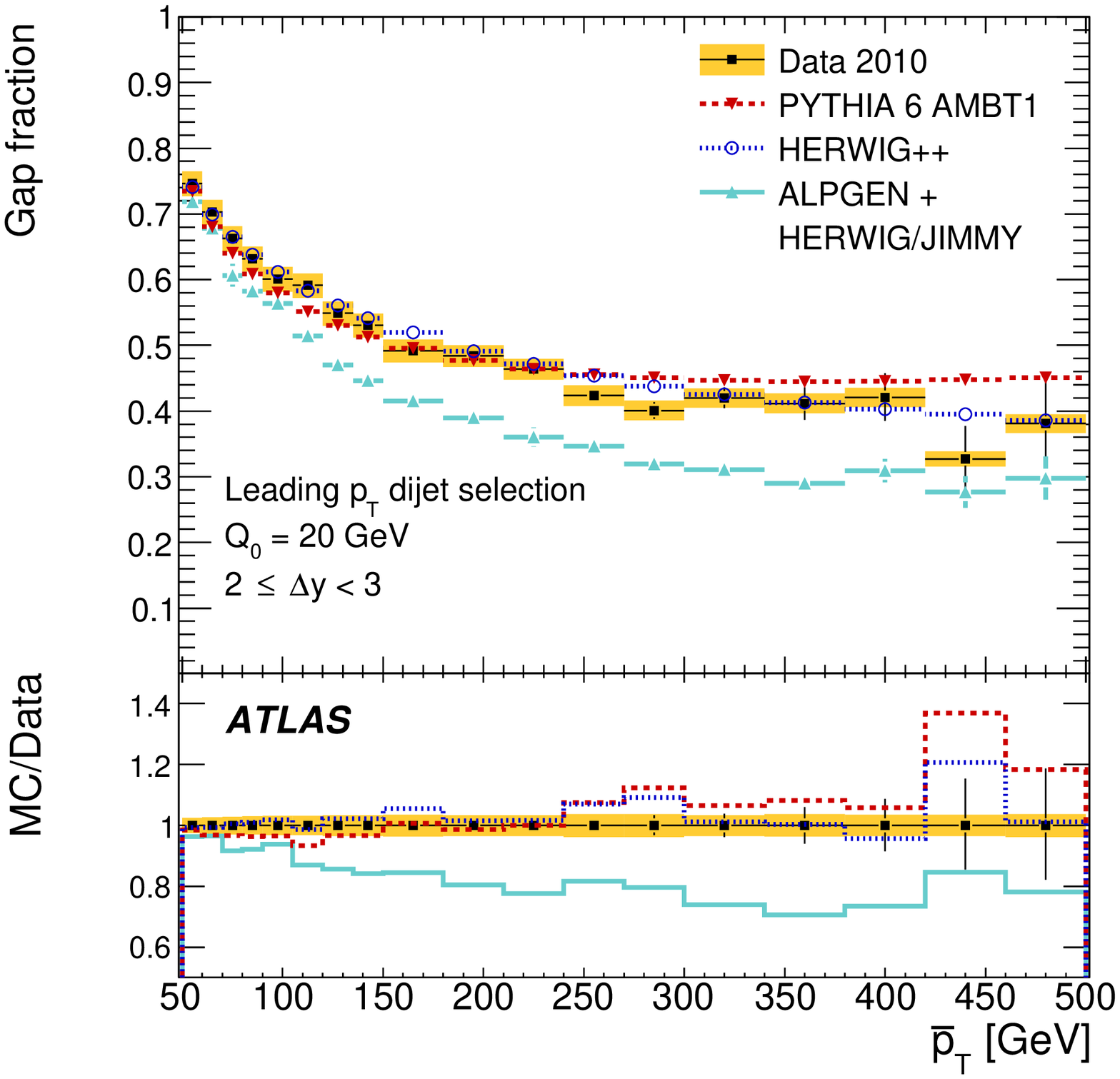}
 }
 
 \caption{Gap fraction as a function of $\Delta y$, given that the dijet system is defined as the leading-$p_{\rm T}^{}$ jets in the event and satisfies $90 \leq \overline{p}_{\rm T}<120$~GeV (a). Gap fraction as a function of $\overline{p}_{\rm T}$ given that the rapidity interval is $2 \leq \Delta y<3$ (b). The (corrected) data are the black points, with error bars representing the statistical uncertainty. The total systematic uncertainty on the measurement is represented by the solid (yellow) band. The dashed (red) points represents the \pythia{} prediction, the dot-dashed (blue) points represents the \herwigpp{} prediction and the solid (cyan) points represents the \alpgen{} prediction.  \label{fig:1}}
\end{figure}

Figure \ref{fig:1} shows the gap fraction as a function of $\Delta y$ and $\overline{p}_{\rm T}$, with the data, corrected for detector effects, compared to the \pythia{}, \herwigpp{} and \alpgen{} event generators.  Both \pythia{} and \herwigpp{} give a good description of the data as a function of $\overline{p}_{\rm T}$. \pythia{} also gives the best description of the data as a function of $\Delta y$. 
\alpgen{} shows the largest deviation from the data, predicting a gap fraction that is too small at large values of $\Delta y$ or $\overline{p}_{\rm T}$.

\section{therotical predictions}
\label{sec:4}
The theoretical predictions were produced using \hej{}  \cite{Andersen:2010ih} and the \powhegbox{} \cite{Frixione:2007nu}.  

\hej{} is a parton-level event generator that provides an all-order description 
of wide-angle emissions of similar transverse momentum, i.e in the limit of large invariant mass between all particles. The events were generated with the MSTW 2008 next-to-leading order (NLO)  PDF set and the partons were clustered into jets using the anti-$k_{\rm t}$ algorithm with distance parameter $R=0.6$. 

The \powhegbox{} provides a full NLO dijet calculation and is interfaced to \pythia{} or \herwig{} to provide all-order resummation of soft and collinear emissions using the parton shower approximation.  The \powheg{} events were generated using the MSTW 2008 NLO PDF set. These events were passed through both \pythia{} (tune AMBT1) and \herwig{} (tune AUET1) to provide different hadron-level predictions. The difference between these two predictions was found to be larger than the intrinsic uncertainty in the NLO calculation.

\section{Final results}
\label{sec:5}
\begin{figure}[h]

\vspace{-0.3cm} 
 \hspace{2cm} (a) \hspace{3.5cm} (b)\\
\resizebox{1.\columnwidth}{!}{

  \includegraphics{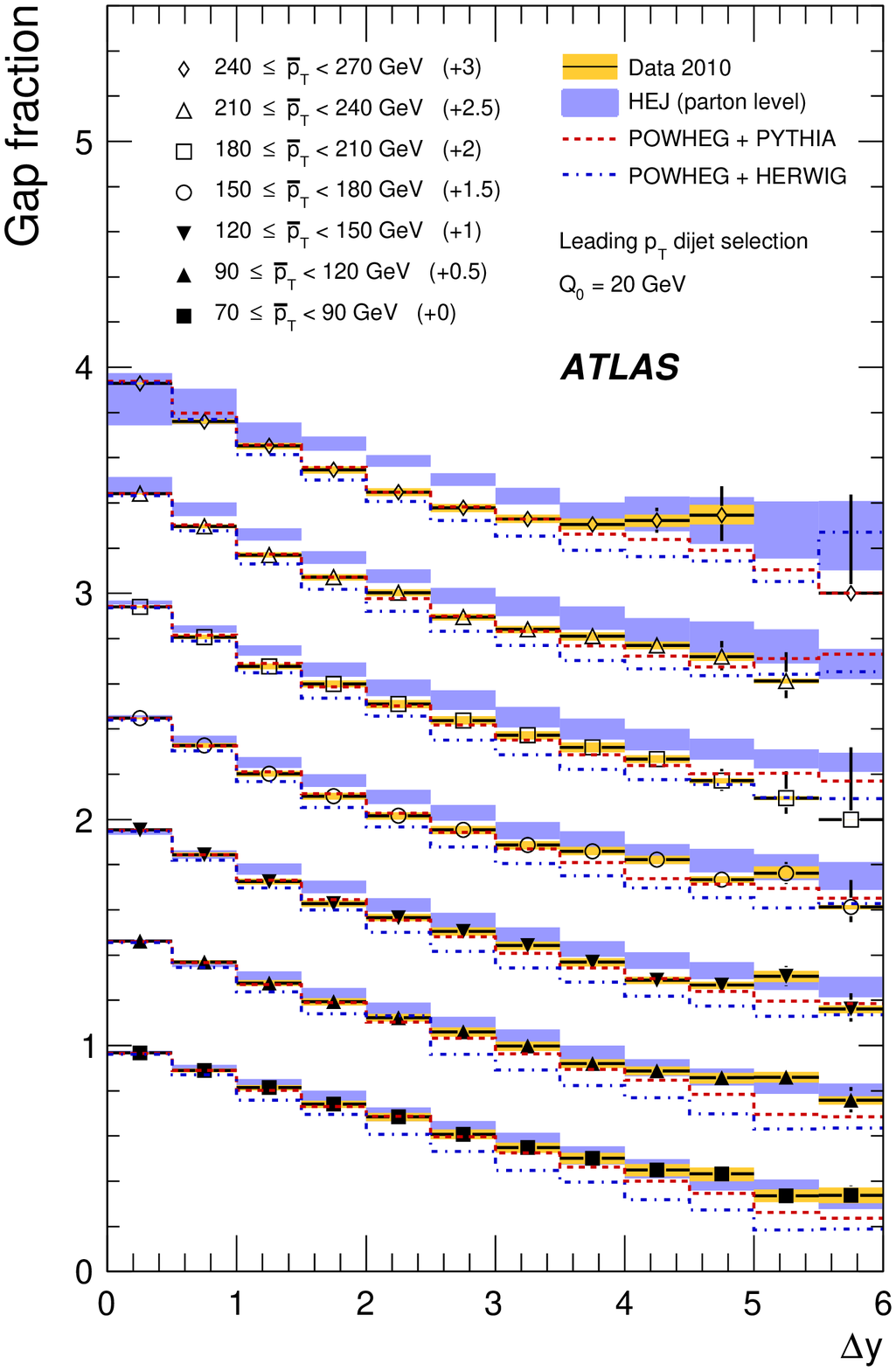} \\	
\includegraphics{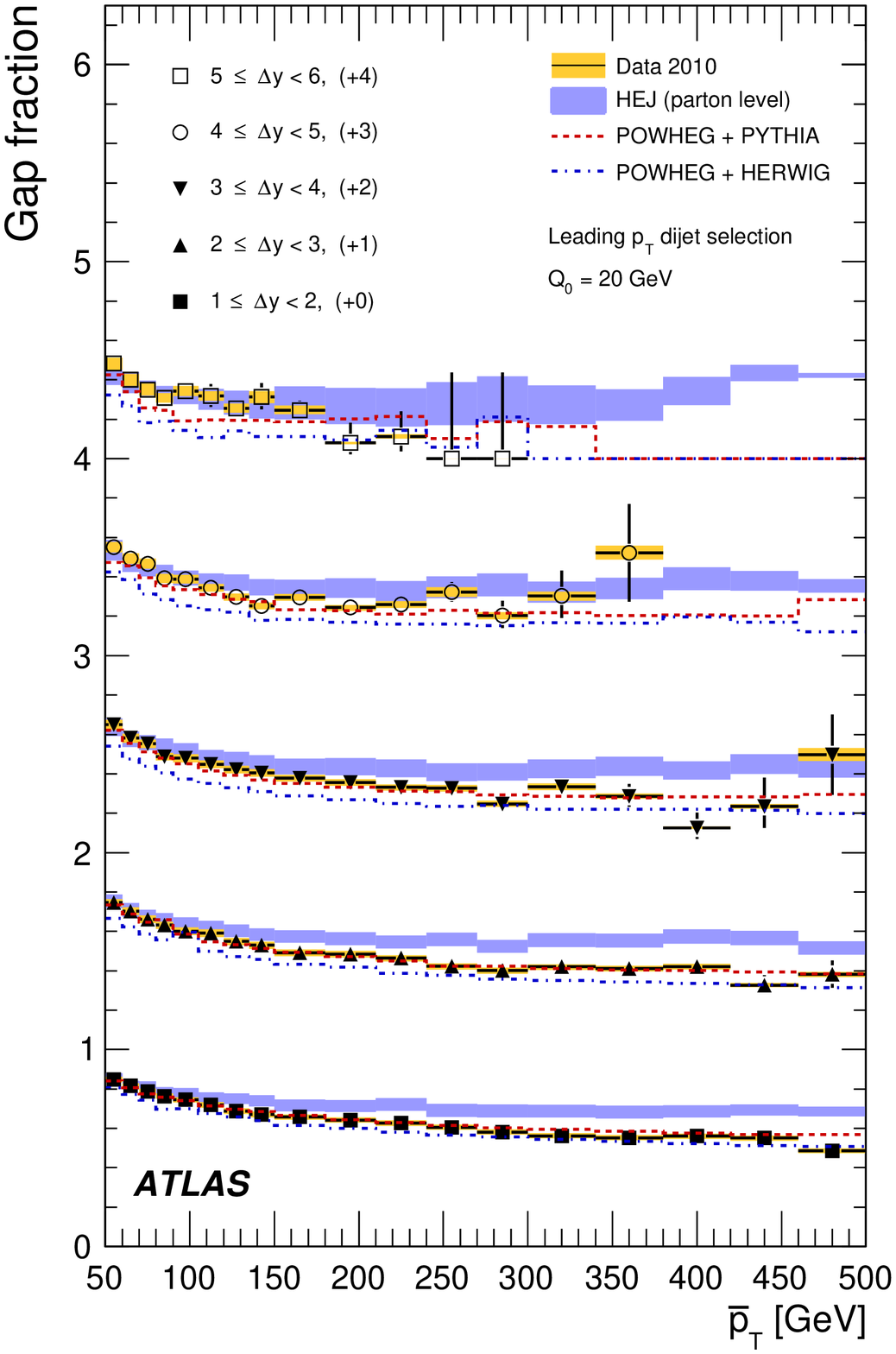}	
 }
\caption{Gap fraction as a function of  $\overline{p}_{\rm T}$ for various $\Delta y$  slices (a) and as a function of   $\Delta y$ for various  $\overline{p}_{\rm T}$  slices (b). The dijet system is defined as the two leading-$p_{\rm T}^{}$ jets in the event. The data are compared to the \hej{} and \powheg{} predictions. The (unfolded) data are the black points, with error bars representing the statistical uncertainty and a solid (yellow) band representing the total systematic uncertainty. The darker (blue) band represents the theoretical uncertainty in the \hej{} calculation from variation of the PDF and renormalisation/factorisation scales. The dashed (red) and dot-dashed (blue) curves represent the \powheg{} predictions after showering, hadronisation and underlying event simulation with \pythia{} and \herwig{}/\jimmy{}, respectively.\label{fig:2}}
\end{figure}

The data are compared to the \hej{} and \powheg{} predictions in Fig.~\ref{fig:2} with the dijet system defined as the two leading-$p_{\rm T}$ jets in the event.  The dependence of the gap fraction on one variable is studied after fixing the phase space of the other variable to well defined and narrow regions. 
The \hej{} prediction describes the data well as a function of $\Delta y$ at low values of $\overline{p}_{\rm T}$. However, at large values of $\overline{p}_{\rm T}$, \hej{} 
predicts too many gap events. This feature is not unexpected as \hej{} calculation is missing higher order QCD effects that become important in this phase space region, i.e those effects that are provided by a traditional parton shower. 
In general, \powhegpyt{} provides the best description of the data. However, at large values of $\Delta y$, the predicted gap fraction deviates from the data. This is expected because the NLO-plus-parton-shower approximation does not contain the contributions to a full QCD calculation that become important as  $\Delta y$ increases. The gap fraction as a function of $\overline{p}_{\rm T}$ is, however, well described by \powhegpyt{} at low $\Delta y$. Furthermore, although the absolute value of the gap fraction is not correct at larger $\Delta y$, the shape of the distributions in $\overline{p}_{\rm T}$ remain well described. 

\begin{figure}
 \hspace{2cm} (a) \hspace{3.5cm} (b)\\
\resizebox{1.\columnwidth}{!}{
  \includegraphics{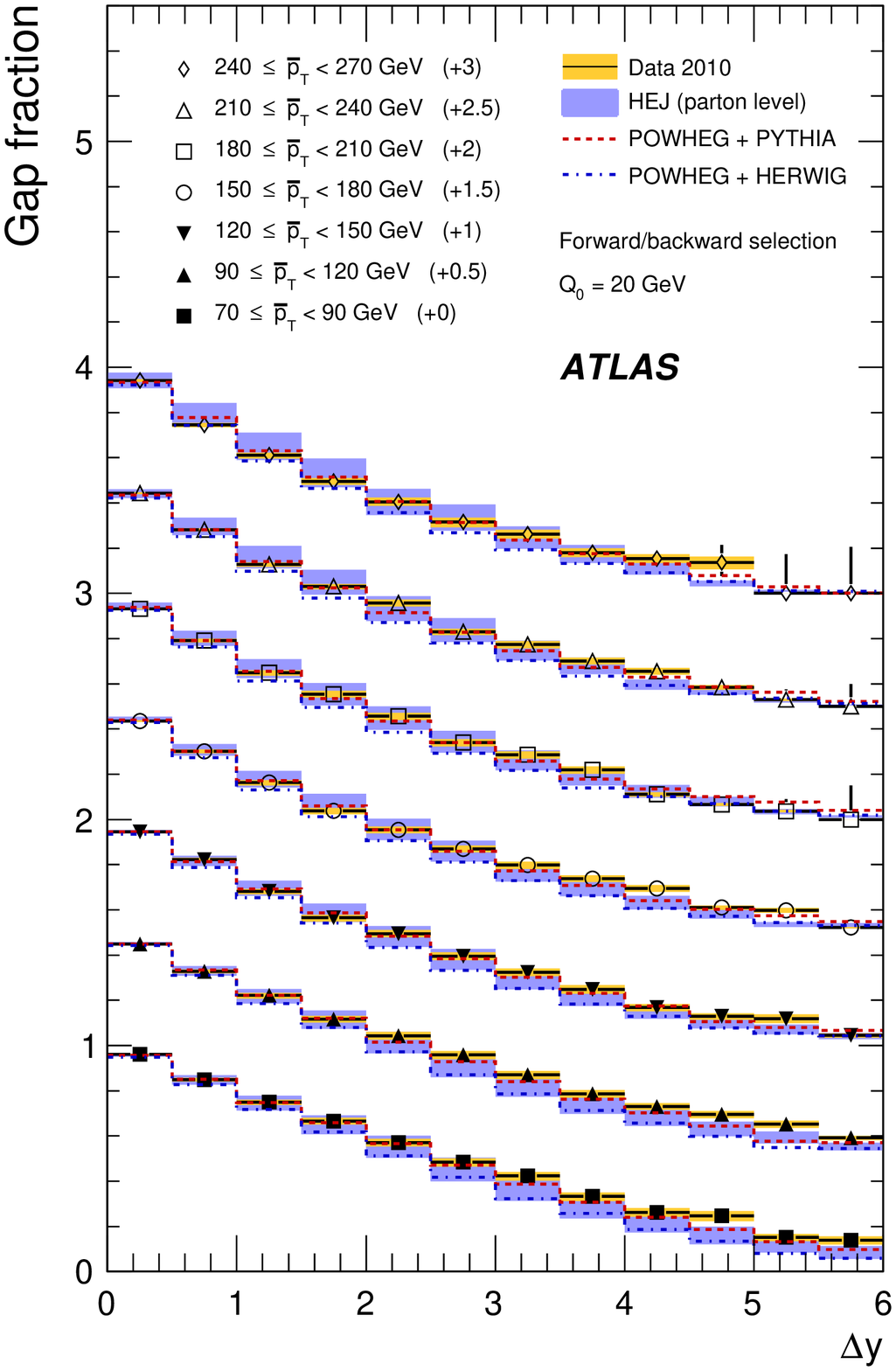} \\	
\includegraphics{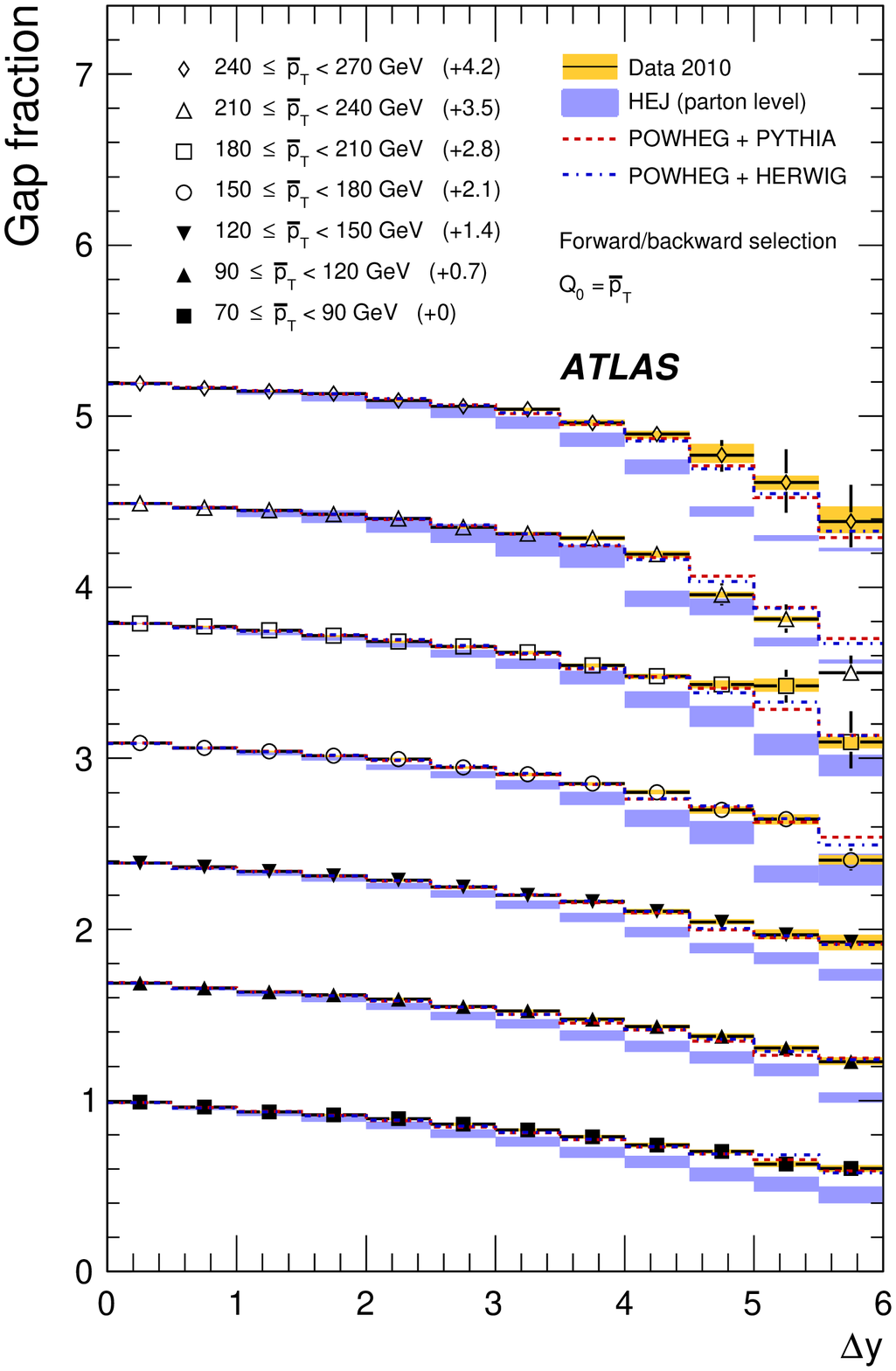}	
 }
\caption{Gap fraction as a function of   $\Delta y$ for various  $\overline{p}_{\rm T}$  slices (b). The dijet system is defined as the most forward and the most backward jets in the event. In (a) the veto scale is set to the default value (20 GeV) while in (b) it is set to $Q_{0} = \overline{p}_{\rm T}$. The data are compared to the \hej{} and \powheg{} predictions. The data and theory are presented in the same way as Fig.~\ref{fig:2}. \label{fig:3}}
\end{figure}

In Fig.~\ref{fig:3} the data are compared to the \hej{} and \powheg{} predictions  with the dijet system defined as  the most forward and backward jets in the event. In Fig.~\ref{fig:3}(a) the veto scale is set to the default value (20 GeV) while in Fig.~\ref{fig:3}(b) it is set to $Q_{0} = \overline{p}_{\rm T}$.  The data are not well described by \hej{} at low values of $\overline{p}_{\rm T}$, implying that the resummation of soft emissions are important for this configuration.  The \powheg{} prediction is similar to the \hej{} prediction in all regions of phase space, that is, both calculations result in a gap fraction that is too small at large $\Delta y$.
In the case where the veto scale is set to $Q_{0} = \overline{p}_{\rm T}$ both \powhegpyt{} and \powhegher{} give a good description of the gap fraction as a function of $\Delta y$, implying a smaller dependence on the generator modelling of the parton shower, hadronisation and underlying event. The \hej{} description of the data, however, does not improve with the increase in veto scale. 

\section{Conclusion}
\label{conc}

A central jet veto was used to study the fraction of events that do not contain hadronic activity in
 the rapidity interval bounded by a dijet system (gap fraction). Data were compared to diverse LO  Monte Carlo event generators and NLO predictions at large rapidity difference and mean transverse momentum of the jets. In most of the phase-space regions presented, the experimental uncertainty is smaller than the theoretical uncertainty.  Furthermore, the experimental uncertainty is much smaller than the spread of LO
 predictions. This measurement of the QCD radiation between widely separated jets can be used to constrain the Monte Carlo simulations for measurements that are sensitive to higher order QCD emissions.



\end{document}